\documentclass{moriond}

\usepackage{amsmath}
\usepackage{xcolor}
\usepackage{subcaption}

\def\be{\begin{equation}}
\def\ee{\end{equation}}
\def\bea{\begin{eqnarray}}
\def\eea{\end{eqnarray}}

\newcommand{\Photo}{\includegraphics[height=45mm]{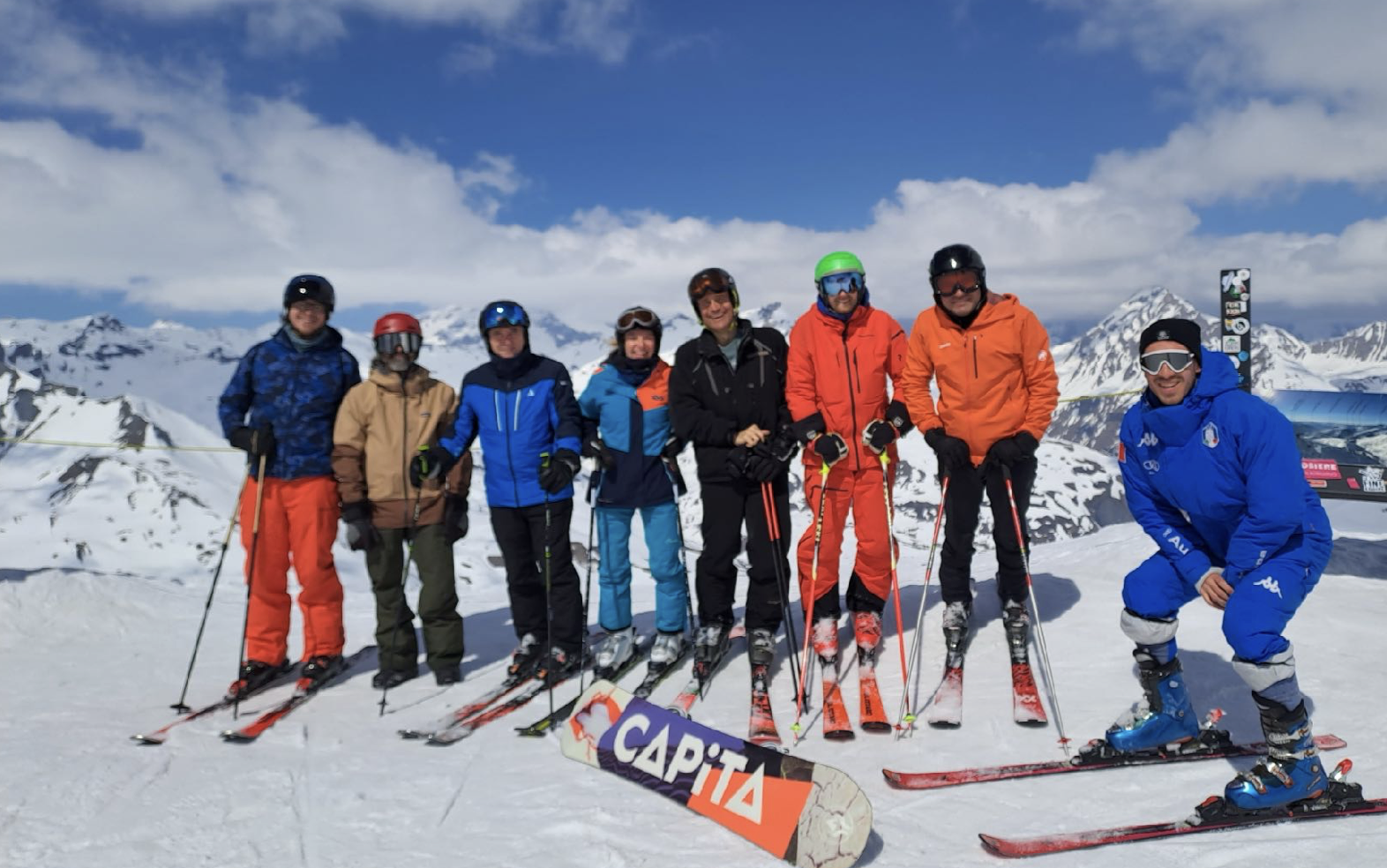}}

\begin{document}
\vspace*{3cm}
\title{Quark mass corrections in di-Higgs production amplitude at high-energy}

\author{ Sebastian Jaskiewicz }

\address{Albert Einstein Center for Fundamental Physics,
Institut f\"ur Theoretische Physik,\\ Universit\"at Bern,
Sidlerstrasse 5, CH-3012 Bern, Switzerland }

\maketitle\abstracts{
A large theoretical uncertainty due to the choice of the top-quark
mass renormalisation scheme is present in 
QCD predictions for Higgs boson pair production. 
In these proceedings, we report on the recent 
progress in tackling these uncertainties 
for the $gg\to HH$ amplitude in the 
high-energy limit $s,|t|,|u| \gg m_t^2 \gg m_H^2$.
Using the Method of Regions and Soft-Collinear Effective Theory,
the leading power in $m_t$ behaviour of the amplitude is 
understood to all orders in the strong coupling 
expansion, and leading logarithmic resummation leads to a significant reduction in the scheme choice uncertainty in the virtual amplitude for
di-Higgs production at high energies. }
\vspace{-1.0cm}
\section{Introduction}
\label{sec:Introduction}

One of the objectives for the High Luminosity phase of the LHC is a measurement of the Higgs self-coupling~\cite{Cepeda:2019klc}.
Precise theoretical predictions are needed for the success of this program~\cite{ATLAS:2025bsu}. Di-Higgs production in gluon fusion is the dominant channel for this process. Sample leading order diagrams 
are depicted in Fig.~\ref{fig:gghh_lo}.
As it has been pointed out in the literature, currently the 
largest source of theoretical uncertainty originates from  
the dependence of the corresponding cross section on the 
choice of the top-quark mass renormalisation scheme~\cite{Baglio:2020ini}. Higher order perturbative calculations will reduce this uncertainty, see e.g. partial NNLO results~\cite{Davies:2025ghl}. 
However, higher order calculations that keep the dependence on all the masses and scales in the process are challenging. Therefore, it is interesting to study whether progress can be made by considering the all-order structure of the top-quark mass corrections in specific kinematic limits of phase-space. 
\vspace{-0.4cm}

\section{\texorpdfstring{$gg\to HH$}{} amplitude at high energies}
\begin{figure}
\begin{center}
\includegraphics[width=0.85\textwidth]{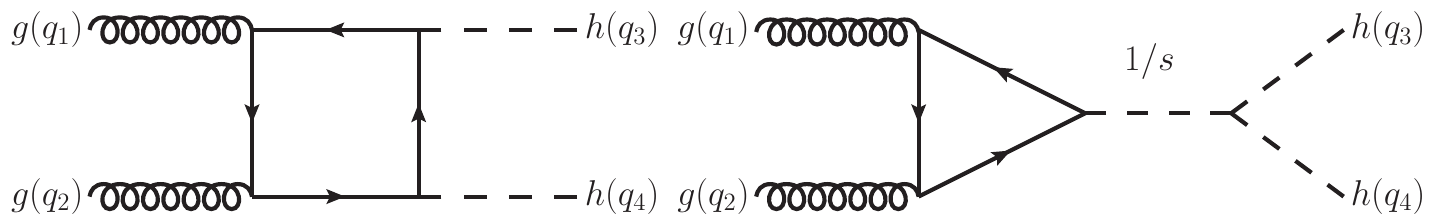}
\caption{\label{fig:gghh_lo} The QCD diagrams contributing to the $gg \to HH$ process at leading order. 
The box diagram on the left-hand side is proportional to $y_t^2$, whereas the triangle diagram on the right-hand side is proportional to $y_t$.}
\end{center}
\end{figure}

First systematic attempt to mitigate the dominant uncertainties stemming from the choice of the top-quark mass renormalisation scheme
has been carried out in a recent work~\cite{Jaskiewicz:2024xkd}.
The study performed in this article focused on investigating the 
behaviour of the $gg\to HH$ virtual amplitude in the high-energy limit. Namely, the following scale hierarchy is imposed 
\begin{equation}\label{eq:high-energy}
s, |t|, |u| \gg m_t^2 \gg m_H^2,
\end{equation}
where $s$, $t$, and $u$ are the usual Mandelstam invariants 
formed in two-to-two particle scattering kinematics, and the $m_t$ and $ m_H$ are the top and Higgs particle masses, respectively. 
The $gg\rightarrow~HH$~amplitude is decomposed in terms of two form factors, which can be chosen to be the helicity amplitudes, $A_1 = -\mathcal{M}^{++} = -\mathcal{M}^{--}$, and $A_2 = -\mathcal{M}^{+-} = -\mathcal{M}^{-+}$. These can be further decomposed: 
\begin{align}
&A_1 = T_F \frac{G_F}{\sqrt{2}} \frac{\alpha_s}{2 \pi} s\  \left[ \frac{3 m_H^2}{s - m_H^2} A_{1,y_t \lambda_3} + A_{1, y_t^2} \right],&
&A_2 = T_F \frac{G_F}{\sqrt{2}} \frac{\alpha_s}{2 \pi} s\  \left[ A_{2, y_t^2} \right],&
\label{eq:Ai}
\end{align}
where  $A_{i,y_t^2}$ are the so-called \textit{box} diagrams, in which both of the Higgs bosons couple to a massive quark line, and 
$A_{i,y_t \lambda_3}$ are the \textit{triangle} diagrams, in which a single off-shell Higgs boson couples to the massive quark line.
In the kinematic set-up described by Eq.~\eqref{eq:high-energy}, the contributions due to the triangle type diagrams on the right-hand side of Fig.~\ref{fig:gghh_lo} are power suppressed, and the dominant contribution arises from the box type diagrams, appearing in the left-hand side diagram of Fig.~\ref{fig:gghh_lo}.
In what follows, we expand the form factors in powers of the strong coupling as 
$
A_{i,j} = \sum_{k=0} \left(\frac{\alpha_s}{2\pi} \right)^k A_{i,j}^{(k)},$
where $ i=1,2,\ j=y_t^2, y_t \lambda_3.
$

The starting point of this investigation~\cite{Jaskiewicz:2024xkd} is
the observation made in literature~\cite{Baglio:2020ini} that in the high-energy limit the
result for one- and two-loop box-type contributions in
the on-shell~($\mathrm{OS}$) renormalisation scheme can be written as~\cite{Baglio:2020ini,Davies:2018qvx,Baglio:2018lrj}
\begin{align}
\label{eq:leading-order}
A^{(0)}_{i,y_t^2} &= y_t^2\, f_i(s,t) 
+ \mathcal{O}(y_t^2 m_t^2)\,, \\ \label{eq:next-to-leading-order}
A^{(1)}_{i,y_t^2} &= 3 C_F \,A^{(0)}_i\,
\log\left[\frac{m_t^2}{s} \right]+ y_t^2 \, g_i(s,t) 
+ \mathcal{O}(y_t^2 m_t^2)\,,
\end{align}
where the $f_i$ functions  are the leading order expressions
for the form factors expanded in $m_t$ to leading power. 
Their explicit form can be found in literature~\cite{Davies:2018qvx}, e.g. $f_1 = {8}/{s}$. The function $g_i(s,t)$ does not contain dependence on $m_t$. 
Interestingly, the leading term proportional to the $C_F$ colour factor at two-loops, i.e. the term displayed in Eq.~\eqref{eq:next-to-leading-order}, originates from the top-quark mass renormalisation counter-term. Consequently,
if we convert the top-quark mass renormalisation scheme to  $\overline{{\rm{MS}}}$, the logarithm appearing in  Eq.~\eqref{eq:next-to-leading-order} instead contains the argument $\mu_t^2/s$. In other words, it has been observed that  in the case of di-Higgs production in gluon-gluon scattering at high-energies,  the logarithmic dependence of the leading power contribution at NLO  is completely determined by the choice of the top-quark mass renormalisation scheme~\cite{Baglio:2020ini}.
In the language of  the Method of Regions (MoR)~\cite{Beneke:1997zp,Smirnov:1990rz,Smirnov:1994tg,Smirnov:2002pj}, this corresponds to the fact that only the  hard region contributes to the leading power amplitude at this order. 
The immediate questions arising from this discussion
addressed by our work~\cite{Jaskiewicz:2024xkd} are
\begin{enumerate}
    \item Why does the box amplitude have such a simple structure?
    \item Is this structure also realised at higher orders, or can super-leading or power-enhanced terms enter in the integrals at higher loop orders to disturb this simple picture?
\end{enumerate}
We have found that since the leading mass-dependent logarithms originate only from the mass renormalisation counter-term, the leading behaviour to all orders can be predicted from the leading order $A^{(0)}_{i,y_t^2}$ result and the top-quark mass running, as well as the universal contributions due to IR matching (massification)~\cite{Jaskiewicz:2024xkd}. Our analysis
is summarised in the next section.

\section{Method of Regions and SCET}
In the first part of the study, we investigate the modes and
regions contributing to the $gg\to HH$ process at fixed orders  using existing tools~\cite{Jantzen:2012mw,Heinrich:2021dbf}. 
We start by considering top-level scalar integrals, which do not contain only the hard region at leading power.  
For example, at the leading order the collinear regions 
enter as well.
At higher loop orders, new regions also appear, as expected \cite{Ma:2023hrt}, and some are even power enhanced at the scalar loop integral level. However, when considering the full amplitude structure, we find that (up to massification corrections)
the complete leading power amplitude is captured by the hard region, i.e.  other regions turn out to give power suppressed contributions to the amplitude.  

\begin{figure}
     \centering
     \begin{subfigure}[b]{0.49\textwidth}
         \centering
         \includegraphics[width=\textwidth]{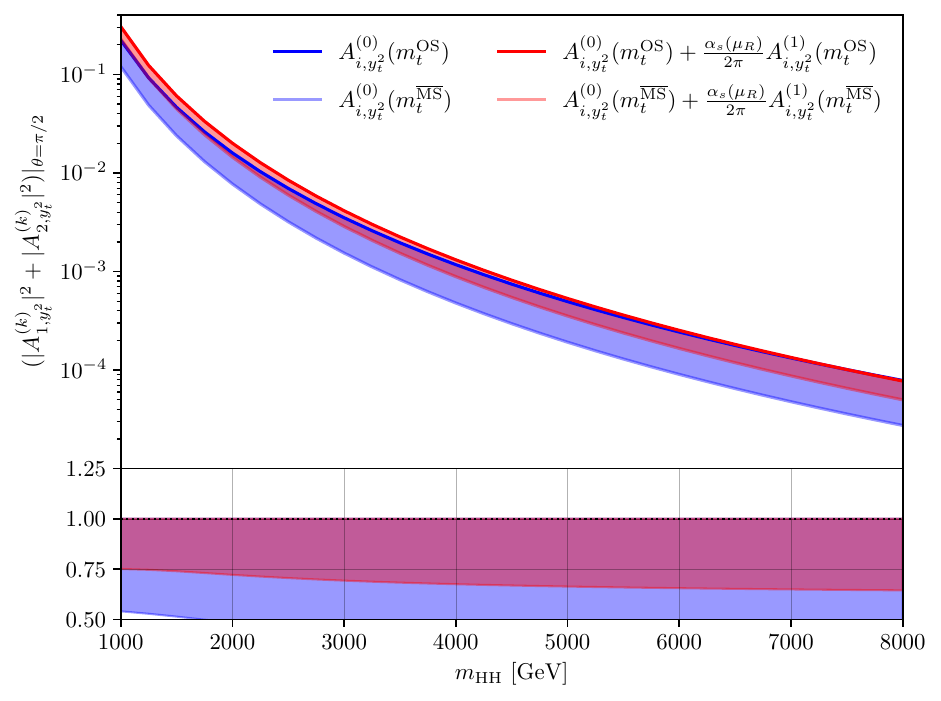}
         \label{sfig:all_sq_msbar_vs_os}
     \end{subfigure}
     \hfill
     \begin{subfigure}[b]{0.49\textwidth}
         \centering
         \includegraphics[width=\textwidth]{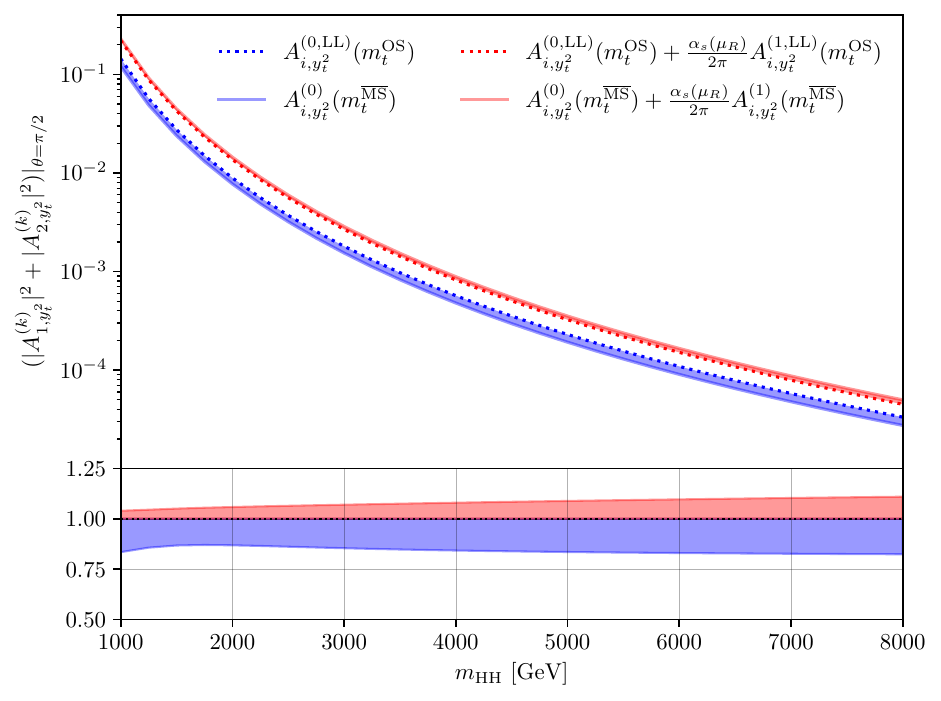}
         \label{sfig:all_sq_msbar_vs_osll}
     \end{subfigure}
        \caption{Comparison of the sum of squared form factors at LO and NLO, where in the left panel the renormalisation of the top-quark mass is performed in $\overline{\mathrm{MS}}$
        and ${\mathrm{OS}}$, and in the right panel, the ${\mathrm{OS}}$
        result supplemented by the resummed tower of LP LLs. 
        We observe significant reduction in the size of the uncertainty due to the choice of mass renormalisation scheme. }
        \label{fig:all_sq_msbar_vs_osll}
\end{figure}

In the second part of the study, we employ effective field theory techniques to formalise the findings from the MoR analysis to all orders in perturbation theory. We use Soft-Collinear Effective Field Theory (SCET) \cite{Bauer:2000yr,Bauer:2001yt,Bauer:2002nz,Beneke:2002ph,Beneke:2002ni},
which captures the relevant degrees of freedom to describe 
the high-energy structure of the $gg\to HH$ amplitude. 
We find that at leading order, the currents that give rise to leading power collinear contributions do not have support for matrix elements with an external gluon or Higgs boson.
Since helicity is conserved in the limit $m_t \to 0$, this result
holds to all orders in $\alpha_s$ and we can conclude that this type 
of contribution will not start to enter the leading power amplitude at 
higher orders. Soft regions involve soft quark contributions which are also power suppressed. Therefore, at the leading logarithmic level at leading power
only the hard region contributes to the $gg\to HH$ amplitude in the
high-energy limit. This analysis answers the questions posed above, 
and leads us to conclude that the leading large logarithms in $m_t^2/s$ appearing
in the amplitude in the OS scheme originate only from the
renormalisation of the top-quark mass. These terms drive the discrepancy in the description of the
amplitude in the OS and $\overline{{\rm{MS}}}$ schemes due to the fact that despite being known to all orders, the $\mu$ dependent part of the conversion between the two schemes $z_m(\mu)$, defined through
\begin{eqnarray}
    \frac{m(\mu)}{M} = 
    \frac{Z^{ \mathrm{OS} }_m}{Z^{\overline{\mathrm{MS}}}_m}
    \equiv z_m(\mu) \,,
\end{eqnarray}
is retained only at the fixed order.  Keeping these known large logarithms at leading-power 
leading-logarithmic level in the conversion factor, using 
\begin{align}\label{eq:schmeme-conv-all-order}
    &m^\text{LL}(\mu) = M \exp \left[ a_{\gamma_m}^\text{LL}(\mu)\right]\, z_m(M),&
    &a_{\gamma_m}^\text{LL}(\mu) = \frac{3 C_F}{2 \beta_0} \ln \left(1 - \frac{\alpha_s(\mu)}{2 \pi} \beta_0 \ln \left(\frac{\mu^2}{M^2} \right) \right),&
\end{align}
 reduces 
the mass scheme dependence of the amplitude at high energy, as can be seen 
in Fig.~\ref{fig:all_sq_msbar_vs_osll}.

\section{Concluding remarks and outlook}
The leading power structure of the $gg\to HH$ amplitude in the high-energy limit has been worked out to all orders in the strong coupling \cite{Jaskiewicz:2024xkd}. The leading large logarithms in $m_t^2/s$ arise
in the scheme conversion factor due to top-quark mass renormalisation. Including these logarithms in the conversion between schemes reduces the mass scheme uncertainties significantly \cite{Jaskiewicz:2024xkd}. 
In the future, it is important to extend the analysis we performed
at amplitude level to physical observables and also beyond the high-energy limit. In particular, it will be important to develop our handle on mass uncertainties 
near the peak of the invariant mass distribution,
i.e. the region $300 \mathrm{GeV} < \sqrt{s} < 800 \mathrm{GeV}$. One possible avenue is to extend the region of validity of the high-energy framework by including subleading power corrections, for which the basis has been developed in~\cite{Jaskiewicz:2024xkd}. However, the factorisation structure at subleading powers is richer than the leading power counterpart, requiring for instance to deal with endpoint divergences~\cite{Beneke:2019oqx,Liu:2019oav,Beneke:2020ibj,Liu:2020wbn,Bell:2022ott,Beneke:2022obx,Liu:2022ajh}. 
Alternatively, it is interesting to develop an all-order handle on the structure of top-quark mass corrections in $gg\to HH$ in the heavy-top limit, the threshold expansion at $s \sim 4 m_t^2$, or the small-$p_T$ expansion.

\section*{Acknowledgments}
\vspace{-0.3cm}
I would like to thank S. Jones, R. Szafron, and Y. Ulrich for collaboration, reading of the manuscript and suggestions. 
This work has been supported by the STFC under grant number ST/X003167/1, the Royal Society University Research Fellowship (URF/R1/201268), and United States Department of Energy under Grant Contract DE-SC0012704.

\vspace{-0.50cm}

\section*{References}

\vspace{-0.3cm}

\end{document}